\documentclass[a4paper,10pt,twoside]{cpc-hepnp}
\usepackage{CJK,upgreek,fancyhdr}
\usepackage{multicol}
\usepackage{graphicx}
\usepackage{booktabs}
\usepackage{amssymb,bm,mathrsfs,bbm,amscd}
\usepackage[tbtags]{amsmath}
\usepackage{lastpage}
\usepackage{hyperref}
\hypersetup{colorlinks=true, citecolor=blue, linkcolor=blue,filecolor=black,urlcolor=blue}

\begin{document}
\begin{CJK*}{GB}{gbsn}

\fancyhead[c]{\small Chinese Physics C~~~Vol. xx, No. x (201x) xxxxxx}
\fancyfoot[C]{\small 010201-\thepage}

\footnotetext[0]{Received 1 September 2017}

\title{Multiplicity fluctuation and correlation of mesons and baryons in ultra-relativistic heavy-ion collisions at LHC \thanks{Supported by National Natural Science Foundation of China (11575100) }}

\author{%
Hai-hong Li(Àºê)$^{1,2}$ 
\quad Feng-lan Shao(ÉÛ·ïÀ¼)$^{1;1)}$\email{shaofl@mail.sdu.edu.cn}%
    \quad Jun Song (Ëξü)$^{2;2)}$\email{songjun2011@jnxy.edu.cn}
}
\maketitle

\address{%
$^1$ School of Physics and Engineering, Qufu Normal University, Shandong 273165, China\\
$^2$  Department of Physics, Jining University, Shandong 273155, China\\
}

\begin{abstract}
    We study the multiplicity fluctuation and correlation of identified mesons and baryons formed at the hadronization by the quark combination mechanism in the context of ultra-relativistic heavy-ion collisions.  Based on the statistical method of free quark combination, we derive the two-hadron multiplicity correlations such as meson-meson and meson-baryon correlations, and take the effects of quark number fluctuation at hadronization into account by a Taylor expansion method. After including the decay contributions, we calculate the dynamical fluctuation observable $\nu_{dyn}$ for $\text{K}\pi$, $p\pi$ and $\text{K}p$ pairs and discuss what underlying physics can be obtained by comparing with the data in Pb-Pb collisions at $\sqrt{s_{NN}}=2.76$ TeV and the simulations from HIJING and AMPT event generators.  
\end{abstract}



\footnotetext[0]{\hspace*{-3mm}\raisebox{0.3ex}{$\scriptstyle\copyright$}2013
Chinese Physical Society and the Institute of High Energy Physics
of the Chinese Academy of Sciences and the Institute
of Modern Physics of the Chinese Academy of Sciences and IOP Publishing Ltd}%

\begin{multicols}{2}


\section{Introduction }

At sufficiently high temperature and/or energy density, strongly interacting matter will undergo a phase transition from hadronic matter to a state in which quarks and gluons are not confined, the quark gluon plasma (QGP) \cite{Shuryak80}. Relativistic heavy-ion collisions serve as the laboratory to experimentally study the properties of QGP \cite{qgp2004}.  Dynamical fluctuations and correlations of (multi-)particle production carry important information on reaction dynamics, and are often used to study the properties of the phase transition between hadronic and partonic matter as well as the QCD critical point \cite{star12Moments,kochFluct,StephanovFluct,Asakawa09,Karsch11,NA492001,Na492009,Na492011,star2009,star2015,koch08,fjh0912,tawfik,Gorenstein2009,torr2007,koch13,koch2010,Konchakovski2009}.  

The multiplicity of produced hadrons is one kind of the most basic quantities of reflecting the reaction dynamics.  The experimental data of event-averaged multiplicity of hadrons not only, by virtue of statistical model, give the information of volume and temperature of the system at chemical freeze-out \cite{becattini02,shm03,Andronc09} but also reveal the microscopic hadronization mechanism of the bulk quark system \cite{ShaoCN09,SJ13}. Fluctuations of hadron multiplicities, and in particular, these of multiplicity ratios carry  the sophisticated information of the dynamical properties of the hot quark matter, and (in particular) the information of the confinement phase transition \cite{koch08,Voloshin02,Christiansen09}.  Many measurements of event-by-event particle ratio fluctuations have been carried out by NA49 collaboration in Pb-Pb collisions at the CERN Super Proton Synchrotron (SPS) \cite{NA492001,Na492009,Na492011}, and by the STAR collaboration in Au+Au collisions at the BNL Relativistic Heavy Ion Collider (RHIC) \cite{star2009,star2015}, and by the ALICE collaboration in Pb-Pb collisions at CERN Large Hadron Collider (LHC) \cite{alicepKpiFluct}.

We note that the available theoretical explanations of these data are usually based on the thermal/statistical models or based on the direct simulations of popular event generators \cite{fjh0912,tawfik,Gorenstein2009,Konchakovski2009,koch2010,torr2007}.  Explanations and predictions from different models of hadron production at different stages are definitely necessary, which will reveal the underlying physics of the experimental data from different viewpoints. In this paper, we study the multiplicity fluctuation and correlation of mesons and baryons created from the hadronization using a quark (re-)combination mechanism. We focus on the effects of quark combination itself and those of quark number fluctuations at hadronization, and derive the multiplicity fluctuation and correlation of mesons and baryons in quark combination mechanism. Applying our formulas, we calculate a dynamical fluctuation observable $\nu_{dyn}$ \cite{Voloshin02,Christiansen09} and discuss what underlying physics can be obtained by comparing with the data in Pb-Pb collisions at $\sqrt{s_{NN}}=2.76$ TeV and the simulations from HIJING and AMPT event generators \cite{alicepKpiFluct}.  Here, we only study the situation of the zero baryon number density, in which the inputs or parameters are relatively few and easily fixed.  Studies of Au+Au collisions at RHIC energies are left to the forthcoming work. 

The paper is organized as follows: Sec.~2 gives a detailed formulation for the inclusive multiplicity of identified hadrons and two-hadron multiplicity correlation in QCM. Sec.~3 shows how to include the effects of quark number fluctuation and correlation at hadronization.  Sec.~4 takes the decay effects in account. Sec.~5 gives the numerical results of a dynamical quantity $\nu_{dyn}$ and discussions of different contributions, and comparison with the experimental data and event generators. Sec.~6 summaries the work.

\section{Hadronic multiplicity and multiplicity correlation in quark combination mechanism }

Quark combination mechanism (QCM) describes the formation of hadrons at hadronization by the combination of quarks and antiquarks neighboring in phase space. The mechanism assumes the effective absence of soft gluon quanta at hadronization and the effective degrees of freedom of QCD matter are quarks and antiquarks with constituent masses at hadronization. Application of quark combination to the bulk quark system produced in relativistic A+A collisions is natural in picture, and QCM has good performance in explaining or reproducing the data of transverse momentum spectra, yields and longitudinal rapidity distributions for various identified hadrons \cite{ShaoCN09,SJ13,biro1999,alcor2000,Fries22003prl,ckm03,hwa04,co2006PRC,sdqcm,fries08review,SJ12FB,wrq12,wrq15}.  In this paper, we do not intend to discuss the space-time details of the combination as those in Refs. \cite{Fries22003prl,ckm03,co2006PRC,fries08review} but concentrate on the multiplicity properties of identified hadrons based on a quark statistical method with the effective constituent quark degrees of freedom.

\subsection{Multiplicity of identified hadrons }

Following the previous work \cite{SFL17}, we write the average multiplicity of identified hadrons after the hadronization of a quark system with given numbers of quarks and antiquarks in the following form
\begin{align}
\overline{N}_{B_{i}} & =N_{B_{i}}^{\left(q\right)}\mathcal{P}_{q_{1}q_{2}q_{3}\rightarrow B_{i}},\label{eqBi}\\
\overline{N}_{M_{i}} & =N_{M_{i}}^{\left(q\right)}\mathcal{P}_{q_{1}\bar{q}_{2}\rightarrow M_{i}},\label{eqMi}
\end{align}
where $N_{B_{i}}^{\left(q\right)}=N_{iter}\prod_{f}\prod_{j=1}^{n_{f,B_{i}}}$$\left(N_{f}-j+1\right)$ is the combination number of three quarks with specific flavors relating to $B_{i}$ formation. $N_{f}$ is the number of quark with flavor $f$ in system. $n_{f,B_{i}}$ is the number of constituent quark $f$ contained in baryon $B_{i}$. $N_{iter}$ is the iteration factor taking to be 1, 3, and 6 for the cases of three identical flavor, two different flavors and three different flavors contained in a baryon, respectively. Examples $N_{p}^{\left(q\right)}=3N_{u}\left(N_{u}-1\right)N_{d}$ and $N_{\Omega^{-}}^{\left(q\right)}=N_{s}\left(N_{s}-1\right)\left(N_{s}-2\right)$ show the evaluation of $N_{B_{i}}^{\left(q\right)}$. $\mathcal{P}_{q_{1}q_{2}q_{3}\rightarrow B_{i}}$ is the combination probability of $q_{1}q_{2}q_{3}\rightarrow B_{i}$.  The meson formula is similar. The combination number of specific-flavor quark antiquark pairs for $M_{i}$ formation is $N_{M_{i}}^{\left(q\right)}=\sum_{k}\omega_{k}\prod_{f}\prod_{j}^{n_{f,M_{i},k}}\left(N_{f}-j+1\right)$ where $f$ runs over all flavors of quarks and antiquarks. This incorporates the case of mixed quark and antiquark flavors for some mesons, e.g.,~$\pi^{0}$ is composed by $u\bar{u}$ and $d\bar{d}$ with weight 1/2, respectively.  Index $k$ runs over all channels of flavor mixing and $\omega_{k}$ is the weight. $n_{f,M_{i},k}$ is the number of constituent (anti-)quark $f$ in $k$ channel, taking to be 1 or 0. $\mathcal{P}_{q_{1}\bar{q}_{2}\rightarrow M_{i}}$ is the combination probability of $q_{1}\bar{q}_{2}\rightarrow M_{i}$.

The combination probabilities $\mathcal{P}_{q_{1}q_{2}q_{3}\rightarrow B_{i}}$ and $P_{q_{1}\bar{q}_{2}\rightarrow M_{i}}$ can be evaluated as
\begin{align}
\mathcal{P}_{q_{1}q_{2}q_{3}\rightarrow B_{i}} & =C_{B_{i}}\frac{\overline{N}_{B}}{N_{qqq}},\label{eqPBi}\\
\mathcal{P}_{q_{1}\bar{q}_{2}\rightarrow M_{i}} & =C_{M_{i}}\frac{\overline{N}_{M}}{N_{q\bar{q}}},\label{eqPMi}
\end{align}
 where $\overline{N}_{B}=\sum_{j}\overline{N}_{B_{j}}$ is the average number of total baryons and $\overline{N}_{M}=\sum_{j}\overline{N}_{M_{j}}$ is total mesons. $N_{q}=\sum_{f}N_{f}$ is total quark number and $N_{\bar{q}}$ total antiquark number. $N_{qqq}=N_{q}\left(N_{q}-1\right)\left(N_{q}-2\right)$ is the total possible number of three quark combinations for baryon formation and $N_{q\bar{q}}=N_{q}N_{\bar{q}}$ is the total possible number of quark antiquark pairs for meson formation. Considering the flavor independence of strong interaction, $\overline{N}_{B}/N_{qqq}$ is used to approximately denote the average probability of three quarks combining into a baryon and $C_{B_{i}}$ is the branch ratio of $B_{i}$ for a given $q_{1}q_{2}q_{3}$ combination. Similarly, $\overline{N}_{M}/N_{q\bar{q}}$ is used to approximately denote the average probability of a quark and antiquark combining into a meson and $C_{M_{i}}$ is the branch ratio to $M_{i}$ for a given flavor $q_{1}\bar{q}_{2}$ combination.

Here we consider only the ground state $J^{P}=0^{-},1^{-}$ mesons and $J^{P}=\left(1/2\right)^{+},\left(3/2\right)^{+}$ baryons in flavor SU(3) group. For mesons
\begin{equation}
C_{M_{i}}=\begin{cases}
\frac{1}{1+R_{V/P}} & \text{for}J^{P}=0^{-}\text{mesons}\\
\frac{R_{V/P}}{1+R_{V/P}} & \text{for}J^{P}=1^{-}\text{mesons},
\end{cases}
\end{equation}
where the parameter $R_{V/P}$ represents the ratios of the $J^{P}=1^{-}$ vector mesons to the $J^{P}=0^{-}$ pseudoscalar mesons of the same flavor composition; for baryons
\begin{equation}
C_{B_{i}}=\begin{cases}
\frac{R_{O/D}}{1+R_{O/D}} & \text{for}J^{P}=\left(\frac{1}{2}\right)^{+}\text{baryons}\\
\frac{1}{1+R_{O/D}} & \text{for}J^{P}=\left(\frac{3}{2}\right)^{+}\text{baryons},
\end{cases}
\end{equation}
except that $C_{\Lambda}=C_{\Sigma^{0}}=R_{O/D}/\left(1+2R_{O/D}\right)$, $C_{\Sigma^{*0}}=1/\left(1+2R_{O/D}\right)$, $C_{\Delta^{++}}=C_{\Delta^{-}}=C_{\Omega^{-}}=1$.  Here $R_{O/D}$ represents the ratios of $J^{p}=\left(1/2\right)^{+}$ octet to the $J^{P}=\left(3/2\right)^{+}$ decuplet baryons of the same flavor composition. \textbf{$R_{V/P}$ }and $R_{O/D}$ are set to be 0.45 and 2.5, respectively.  The number of constituent quarks is conserved at hadronization, which means
\begin{align}
\overline{N}_{M}+3\overline{N}_{B} & =N_{q},\\
\overline{N}_{M}+3\overline{N}_{\bar{B}} & =N_{\bar{q}}.
\end{align}
In Ref. \cite{SJ13}, we have obtained the empirical solution of $\overline{N}_{M}$, $\overline{N}_{B}$ and $\overline{N}_{\bar{B}}$ for the hadronization of large quark system, which was tested against the RHIC data. As the net-baryon number is negligible at LHC energies, the formula of baryon number is simple $\overline{N}_{B}\approx\overline{N}_{\bar{B}}\approx N_{q}/15$ and meson number is obtained by the above quark number conservation. In addition, the conservation of specific quark flavor is also satisfied,
\begin{equation}
\sum_{\alpha}n_{f,\alpha}\overline{N}_{\alpha}=N_{f}.
\end{equation}
Index $\alpha$ denotes the hadron of kind $\alpha$ and $f=u,d,s,\bar{u},\bar{d},\bar{s}$ denotes the flavor of quarks and antiquarks.  

\subsection{Two-hadron multiplicity correlations}
\subsubsection{Two-baryon correlation}
We start from the pair production of two baryons $B_{i}$ and $B_{j}$
\begin{equation}
\overline{N}_{B_{i}B_{j}}=C_{B_{i}}C_{B_{j}}N_{B_{i}B_{j}}^{\left(q\right)}\frac{\overline{N_{B}\left(N_{B}-1\right)}}{N_{6q}}.\label{eqBab}
\end{equation}
Here, $N_{B_{i}B_{j}}^{\left(q\right)}$ is the possible cluster number of six specific quarks relating to two-baryon joint formation, and is evaluated as $N_{B_{i}B_{j}}^{\left(q\right)}=N_{iter,B_i}N_{iter,B_j}\prod_{f}\prod_{j=1}^{n_{f,B_{i}}+n_{f,B_{j}}}$$\left(N_{f}-j+1\right)$ where $f$ runs over all quark flavors. $N_{6q}=\prod_{i=1}^{i=6}\left(N_{q}-i+1\right)$ is the total possible cluster number of six quarks. $\overline{N_{B}\left(N_{B}-1\right)}$ is the number of two-baryon pairs and \textbf{$\overline{N_{B}\left(N_{B}-1\right)}/N_{6q}$ }gives the average probability of six quarks combining into two baryons.  We rewrite the term $C_{B_{i}}C_{B_{j}}N_{B_{i}B_{j}}^{\left(q\right)}/N_{6q}=P_{B_{i}}P_{B_{j}}\left(1-A_{B_{i}B_{j}}\right)$ where $P_{B_{i}}=\overline{N}_{B_{i}}/\overline{N}_{B}$ denotes the fraction of baryon $i$ in total baryon production. $A_{B_{i}B_{j}}$ is a small quantity of the magnitude $\mathcal{O}\left(N_{f}^{-1}\right)$.  Using the relation $\overline{N_{B}\left(N_{B}-1\right)}=\overline{\sigma}_{B}^{2}+\overline{N}_{B}\left(\overline{N}_{B}-1\right)$, we have
\begin{align}
\overline{C}_{B_{i}B_{j}} & =\overline{N_{B_{i}}N_{B_{j}}}-\overline{N}_{B_{i}}\overline{N}_{B_{j}}\nonumber \\
 & =\overline{N}_{B_{i}B_{j}}+\delta_{ij}\overline{N}_{B_{i}}-\overline{N}_{B_{i}}\overline{N}_{B_{j}}\nonumber \\
 & =P_{B_{i}}\overline{N}_{B}\left(\delta_{i,j}-P_{B_{j}}\right) \label{eq:c2b} \\ 
    &   +P_{B_{i}}P_{B_{j}}\left[\left(1-A_{B_{i}B_{j}}\right)\overline{\sigma}_{B}^{2}-A_{B_{i}B_{j}}\overline{N}_{B}\left(\overline{N}_{B}-1\right)\right].\nonumber 
\end{align}

The first term in right-hand side is the result of binomial distribution and is the leading term in the two-baryon multiplicity correlations. The second term in right-hand side is quite small relative to the former.  For more detail discussions and numerical results on the above two-baryon correlation, we refer readers to Ref. \cite{SJ17fluct}. The variance of total baryons $\overline{\sigma}_{B}^{2}$ is not determined analytically at present and we adopt a parameterization $\overline{\sigma}_{B}^{2}\approx0.36\overline{N}_{B}$ according to the simulation of a quark combination model developed by Shandong Group \cite{sdqcm,ShaoCN09}.

For baryon-antibaryon multiplicity correlation, we firstly write the pair production of baryon-antibaryon
\begin{equation}
\overline{N}_{B_{i}\bar{B}_{j}}=C_{B_{i}}C_{\bar{B}_{j}}N_{B_{i}\bar{B}_{j}}^{\left(q\right)}\frac{\overline{N_{B}N_{\bar{B}}}}{N_{3q3\bar{q}}},
\end{equation}
 where $N_{B_{i}\bar{B}_{j}}^{\left(q\right)}$ is the possible cluster number of three quarks and three antiquarks relating to $B_{i}\bar{B}_{j}$ joint formation, and is evaluated as $N_{B_{i}\bar{B}_{j}}^{\left(q\right)}= N_{iter,B_i} N_{iter,\bar{B}_j}\prod_{f}\prod_{j=1}^{n_{f,B_{i}}+n_{f,\bar{B}_{j}}}\left(N_{f}-j+1\right)$ where index $f$ runs overs all flavors of quarks and antiquarks.
$N_{3q3\bar{q}}=N_{qqq}N_{\bar{q}\bar{q}\bar{q}}$ and $\overline{N_{B}N_{\bar{B}}}=\overline{\sigma}_{B}^{2}+\overline{N}_{B}\overline{N}_{\bar{B}}$.
Using the denotation $P_{B_{i}}$and $P_{\bar{B}_{j}},$ we have
\begin{align}
\overline{C}_{B_{i}\bar{B_{j}}} & =\overline{N_{B_{i}}N_{\bar{B}_{j}}}-\overline{N}_{B_{i}}\overline{N}_{\bar{B}_{j}}=P_{B_{i}}P_{\bar{B}_{j}}\overline{\sigma}_{B}^{2}. \label{eq:cBBbar}
\end{align}

\subsubsection{Two-meson correlation}

The pair production of two mesons $M_{i}$ and $M_{j}$ is written as 
\begin{equation}
\overline{N}_{M_{i},M_{j}}=C_{M_{i}}C_{M_{j}}N_{M_{i}M_{j}}^{\left(q\right)}\frac{\overline{N_{M}\left(N_{M}-1\right)}}{N_{2q2\bar{q}}}\label{eq:Mab}
\end{equation}
Here $N_{M_{i}M_{j}}^{\left(q\right)}$ is the possible cluster number of two specific quarks and two specific antiquarks relating to two-meson joint formation, and is evaluated as $\sum_{kl}\omega_{k}\omega_{l}\prod_{f}\prod_{j}^{n_{f,M_{i},k}+n_{f,M_{j},l}}\left(N_{f}-j+1\right)$ where index $f$ runs over all flavors of quarks and antiquarks. $N_{2q2\bar{q}}=N_{q}\left(N_{q}-1\right)N_{\bar{q}}\left(N_{\bar{q}}-1\right)$ is total number of all $qq\bar{q}\bar{q}$ combinations. $\overline{N_{M}\left(N_{M}-1\right)}$ is number of two-meson pair and $\overline{N_{M}\left(N_{M}-1\right)}/N_{2q2\bar{q}}$ is the average probability of two quarks and two antiquarks combining into two mesons. We rewrite the term $C_{M_{i}}C_{M_{j}}\frac{N_{M_{i}M_{j}}^{\left(q\right)}}{N_{2q2\bar{q}}}=P_{M_{i}}P_{M_{j}}(1-A_{M_{i}M_{j}})$ where $P_{M_{i}}=\overline{N}_{M_{i}}/\overline{N}_{M}$ denotes the fraction of meson $i$ in total meson production. $A_{M_{i}M_{j}}$ is a small quantity of the magnitude $\mathcal{O}\left(N_{f}^{-1}\right)$.  Using $\overline{N_{M}\left(N_{M}-1\right)}=\overline{\sigma}_{M}^{2}+\overline{N}_{M}\left(\overline{N}_{M}-1\right)$, we have
\begin{align}
\overline{C}_{M_{i}M_{j}} & =\overline{N_{M_{i}}N_{M_{j}}}-\overline{N}_{M_{i}}\overline{N}_{M_{j}}\nonumber \\
 & =P_{M_{i}}\overline{N}_{M}\left(\delta_{i,j}-P_{M_{j}}\right) \label{eq:c2m} \\
    & +P_{M_{i}}P_{M_{j}}\left[\left(1-A_{M_{i}M_{j}}\right)\overline{\sigma}_{M}^{2}-A_{M_{i}M_{j}}\overline{N}_{M}\left(\overline{N}_{M}-1\right)\right]. \nonumber
\end{align}
This is quite similar to two-baryon correlations in Eq.~(\ref{eq:c2b}).  The first terms in right-hand side is the form of the Binomial distribution.  The second term in right-hand side is related to the fluctuation of global meson production and effects of finite quark numbers by $A_{M_{i}M_{j}}$ coefficient. We notice that the influence of second term on two-meson correlation is relatively obvious for some hadron species.

\subsubsection{Baryon-meson multiplicity correlation}
The similar procedure is applied to baryon-meson correlations, and we firstly write the pair production of a baryon and a meson
\begin{equation}
\overline{N}_{B_{i}M_{j}}=C_{B_{i}}C_{M_{j}}N_{B_{i}M_{j}}^{\left(q\right)}\frac{\overline{N_{B}N_{M}}}{N_{4q\bar{q}}},
\end{equation}
 where $N_{B_{i}M_{j}}^{\left(q\right)}$ is the possible cluster number of four specific quarks and an antiquark relating to $B_{i}M_{j}$ joint formation, and is evaluated as $N_{B_{i}M_{j}}^{\left(q\right)}=N_{iter,B_i}\sum_{k}\omega_{k}\prod_{f}\prod_{j=1}^{n_{f,B_{i}}+n_{f,M_{j},k}}\left(N_{f}-j+1\right)$.  $N_{4q\bar{q}}=N_{q}\left(N_{q}-1\right)\left(N_{q}-2\right)\left(N_{q}-3\right)N_{\bar{q}}$ is total cluster number of four quarks and an antiquark. We rewrite $\overline{N_{B}N_{M}}=-3\overline{\sigma}_{B}^{2}+\overline{N}_{B}\overline{N}_{M}$ using the unitarity $\overline{N}_{M}+3\overline{N}_{B}=N_{q}$ and rewrite the term $C_{B_{i}}C_{M_{j}}\frac{N_{B_{i}M_{j}}^{\left(q\right)}}{N_{4q\bar{q}}}=P_{B_{i}}P_{M_{j}}\left(1-A_{B_{i}M_{j}}\right)$, we obtain
\begin{align}
    &\overline{C}_{B_{i}M_{j}}  =\overline{N_{B_{i}}N_{M_{j}}}-\overline{N}_{B_{i}}\overline{N}_{M_{j}}\nonumber \\
 & =P_{B_{i}}P_{M_{j}}\left(1-A_{B_{i}M_{j}}\right)\left(-3\overline{\sigma}_{B}^{2}+\overline{N}_{B}\overline{N}_{M}\right)-P_{B_{i}}P_{M_{j}}\overline{N}_{B}\overline{N}_{M}\nonumber \\
 & =-P_{B_{i}}P_{M_{j}}\left[3\left(1-A_{B_{i}M_{j}}\right)\overline{\sigma}_{B}^{2}+A_{B_{i}M_{j}}\overline{N}_{B}\overline{N}_{M}\right].\label{eq:cbm}
\end{align}
The antibaryon-meson correlation is obtained by taking the charge conjugation transformation on the Eq.~(\ref{eq:cbm}).

\section{Effects of quark number fluctuation and correlation at hadronization \label{sec:qfnc}}

The produced quark system in heavy ion collisions is always varied in size event-by-event, and we should take the effects of the fluctuation of quark numbers into account. Suppose the number of quarks and that of antiquarks follow a distribution $\mathcal{P}\left(\left\{ N_{f}N_{\bar{f}}\right\} ;\left\{ \langle N_{f}\rangle,\langle N_{\bar{f}}\rangle\right\} \right)$ around the event average $\langle N_{f}\rangle$ and $\langle N_{\bar{f}}\rangle$ with $f=u,d,s$, the event average of a hadronic quantity $A_{h}$ is
\begin{equation}
\langle A_{h}\rangle=\sum_{\left\{ N_{f}N_{\bar{f}}\right\} }\overline{A}_{h}\mathcal{P}\left(\left\{ N_{f}N_{\bar{f}}\right\} ;\left\{ \langle N_{f}\rangle,\langle N_{\bar{f}}\rangle\right\} \right),\label{eq:AveAh}
\end{equation}
 where $\overline{A}_{h}$ is the result at given quark numbers and antiquark numbers. We expand the $\overline{A}_{h}$ as the Taylor series at the event average of quark numbers $\left\{ \langle N_{f}\rangle,\langle N_{\bar{f}}\rangle\right\},$
\begin{align}
\overline{A}_{h} & =\left.\overline{A}_{h}\right|_{\langle\cdot\rangle}+\sum_{f_{1}}\left.\frac{\partial\overline{A}_{h}}{\partial N_{f_{1}}}\right|_{\langle\cdot\rangle}\delta N_{f_{1}}\nonumber \\
 & +\frac{1}{2}\sum_{f_{1}f_{2}}\left.\frac{\partial^{2}\overline{A}_{h}}{\partial N_{f_{1}}\partial N_{f_{2}}}\right|_{\langle\cdot\rangle}\delta N_{f_{1}}\delta N_{f_{2}}+\mathcal{O}\left(N_{f}^{-2}\right),
\end{align}
where indexes $f_{1}$ and $f_{2}$ run over all flavors of quarks and antiquarks and $\delta N_{f_{1}}=N_{f_{1}}-\langle N_{f_{1}}\rangle$.  Subscript $\langle\cdot\rangle$ denotes the evaluation at event average.  Substituting the above into Eq.~(\ref{eq:AveAh}), we get 
\begin{equation}
\langle A_{h}\rangle=\overline{A}_{h}+\frac{1}{2}\sum_{f_{1}f_{2}}\frac{\partial^{2}\overline{A}_{h}}{\partial N_{f_{1}}\partial N_{f_{2}}}C_{f_{1}f_{2}}+\mathcal{O}\left(\langle N_{f}\rangle^{-2}\right).
\end{equation}
where $C_{f_{1}f_{2}}=\langle\delta N_{f_{1}}\delta N_{f_{2}}\rangle$ is two-body correlation function of quarks and antiquarks and we drop the subscript $\langle\cdot\rangle$ for short. Applying it to the multiplicity quantities $N_{\alpha}$ and $N_{\alpha}N_{\beta}$, we have, up to second order
\begin{align}
\langle N_{\alpha}\rangle & =\overline{N}_{\alpha}+\frac{1}{2}\sum_{f_{1}f_{2}}\left(\partial_{12}\overline{N}_{\alpha}\right)C_{f_{1}f_{2}},\label{eq:qnfc_Na}\\
C_{\alpha\beta} & =\overline{C}_{\alpha\beta}+\frac{1}{2}\sum_{f_{1}f_{2}}\left[2\partial_{1}\overline{N}_{\alpha}\,\partial_{2}\overline{N}_{\beta}+\partial_{12}\overline{C}_{\alpha\beta}\right]C_{f_{1}f_{2}},\label{eq:qnfc_cab}
\end{align}
where we have used the abbreviation $\partial_{1}\equiv\frac{\partial}{\partial N_{f_{1}}}$ and $\partial_{12}=\frac{\partial^{2}}{\partial N_{f_{1}}\partial N_{f_{2}}}$.  Because multiplicity $\overline{N}_{\alpha}$ is almost homogeneous function of quark numbers and quark correlations $C_{f_{1}f_{2}}$ are usually the magnitude of $\langle N_{f_{1}}\rangle$ or $\langle N_{f_{2}}\rangle$, $\left(\partial_{12}\overline{N}_{\alpha}\right)C_{f_{1}f_{2}}$ has the magnitude of $\overline{N}_{\alpha}/N_{f}$, and therefore affect the $\langle N_{\alpha}\rangle$ at $1/\langle N_{f}\rangle$ level.  It is very small as quark numbers are large. For two-hadron correlations, the effects of quark number fluctuation are the magnitude of $\overline{N}_{\alpha}\overline{N}_{\beta}/N_{q}$ which is comparable with $\overline{C}_{\alpha\beta}$, and therefore will significantly influence two-hadron correlations. 

\begin{center}
\includegraphics[width=\linewidth]{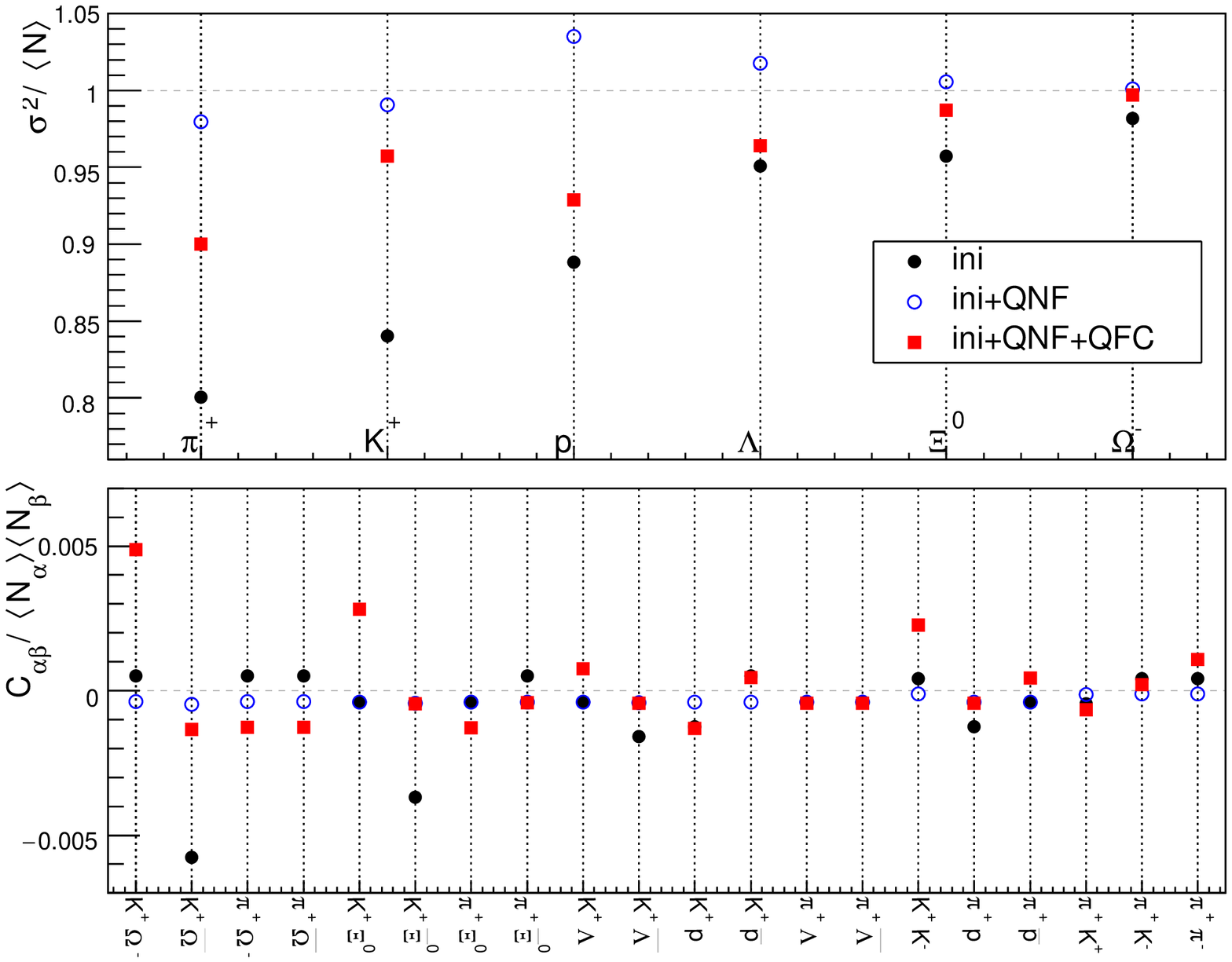}
\figcaption{\label{fig1}   Variance and two-hadron correlation for a few hadrons. Results by combination itself are shown as solid circles (ini); results after considering the quark number fluctuation (QNF) are shown as open circles (ini+QNF), and results after further considering the quark number flavor conservation (QFC) are shown as solid squares (ini+QNF+QFC).  }
\end{center}

As examples, in Fig.~1 we show the numerical results of variance and two-hadron correlation for a few hadrons. Quark numbers are taken $\langle N_{u}\rangle=\langle N_{d}\rangle=\langle N_{\bar{u}}\rangle=\langle N_{\bar{d}}\rangle=1137$ and $\langle N_{s}\rangle=\langle N_{\bar{s}}\rangle=478$, which is corresponding to the size of quark system in the unit rapidity interval in 0-5\% Pb-Pb collisions at $\sqrt{s_{NN}}$=2.76 TeV. Here, results directly by the combination, i.e.,~Eqs. (\ref{eq:c2m}) and (\ref{eq:cbm}), are shown as solid circles. Open circles show results of Eq.~(\ref{eq:qnfc_cab}) after including only the effects of quark number fluctuations $C_{ff}=\langle N_{f}\rangle$. Solid squares show results of Eq.~(\ref{eq:qnfc_cab}) after further including the flavor conservation of quarks and antiquarks $C_{f\bar{f}}=\langle N_{f}\rangle$.  We see that effects of quark number fluctuation and correlation are non-trivial in general and are varied with hadron species. Lattice QCD calculations \cite{Gavai05,Mukherjee06,Dht15} show that off-diagonal flavor susceptibilities of quark numbers in the vicinity of (de-)confinement phase transition region are quite small compared with diagonal ones, $\chi_{us}/\chi_{ss} \approx -0.05$ and $\chi_{ud}/\chi_{uu} \approx -0.05$. Off-diagonal quark correlations $C_{f_{1}f_{2}}$ and $C_{f_{1}\bar{f}_{2}}$ are also usually small and therefore are not discussed here.

\section{Decay effects}

The influence of resonance decay is usually important and complex.  For a resonance $\alpha$, the probability of $k$-th decay channel is denoted as $D_{\alpha k}$ and is taken from PDG\cite{pdg2014}. The number of stable hadron $a$ in this channel is denoted as $\omega_{\alpha k}^{a}$. The number of decay channels for $\alpha$ is denoted as $d_{\alpha}$. The multiplicity of stable hadron $a$ is 
\begin{equation}
\langle N_{a}^{\left(f\right)}\rangle=\sum_{\alpha}\langle N_{\alpha}^{\left(i\right)}\rangle\left(\delta_{\alpha,a}+\left(1-\delta_{\alpha,a}\right)\sum_{k=1}^{d_{\alpha}}D_{\alpha k}\omega_{\alpha k}^{a}\right),\label{eq:Naf}
\end{equation}
where we use the superscript $\left(i\right)$ to denote results for initially produced hadrons and $\left(f\right)$ for final hadrons including decay contributions. For two-hadron correlation,
\end{multicols}
\begin{align}
C_{ab}^{\left(f\right)} & =\sum_{\epsilon\sigma}C_{\epsilon\sigma}^{\left(i\right)}\left[\delta_{\epsilon,a}+\left(1-\delta_{\epsilon,a}\right)\sum_{k=1}^{d_{\epsilon}}D_{\epsilon k}\omega_{\epsilon k}^{a}\right]\left[\delta_{\sigma,b}+\left(1-\delta_{\sigma,b}\right)\sum_{k=1}^{d_{\sigma}}D_{\sigma k}\omega_{\sigma k}^{b}\right]\nonumber \\
 & +\sum_{\epsilon}\langle N_{\epsilon}^{\left(i\right)}\rangle\left(1-\delta_{\epsilon,a}\right)\left(1-\delta_{\epsilon,b}\right)\sum_{k,k'=1}^{d_{\epsilon}}D_{\epsilon k}\left(\delta_{k,k'}-D_{\epsilon k'}\right)\omega_{\epsilon k}^{a}\omega_{\epsilon k'}^{b},\label{eq:Cabf}
\end{align}
 which receives the superposition of two-body correlations of other hadrons which can decay into $a$ and/or $b$. In particular, if a resonance $\epsilon$ can decay into both $a$ and $b$, the $\langle N_{\epsilon}^{\left(i\right)}\rangle$ contribution term arises, second term in right-hand side, with a positive sign if $a$ and $b$ come from the same decay channel ($k=k'$) and a negative sign if $a$ and \textbf{$b$ }come form different decay channels ($k\neq k'$). Extensions of Eqs. (\ref{eq:Naf}) and (\ref{eq:Cabf}) to cascade decays are straightforward but the formulas are too lengthy to be shown in this paper. We use the full decay formulas in practical calculations. 

\begin{multicols}{2}

\section{Dynamical fluctuation $\nu_{dyn}$ of $p\pi$, K$\pi$ and K$p$ pairs}

In this section, we discuss an observable $\nu_{dyn}$ which is the combination of two-hadron correlations and is proposed as an effective probe of the dynamical fluctuations \cite{Voloshin02}. It takes the form $\langle\left(\frac{N_{A}}{\langle N_{A}\rangle}-\frac{N_{B}}{\langle N_{B}\rangle}\right)^{2}\rangle$ and provides a measurement of the dynamical variance for the difference between the relative numbers of the two particle species A and B. Subtracting the base line of purely statistical fluctuation $\frac{1}{\langle N_{A}\rangle}+\frac{1}{\langle N_{B}\rangle}$, the generalized definition of $\nu_{dyn}$ is 
\begin{align}
\nu_{dyn,AB} & =\frac{\langle N_{A}\left(N_{A}-1\right)\rangle}{\langle N_{A}\rangle^{2}}+\frac{\langle N_{B}\left(N_{B}-1\right)\rangle}{\langle N_{B}\rangle^{2}}-2\frac{\langle N_{A}N_{B}\rangle}{\langle N_{A}\rangle\langle N_{B}\rangle}\nonumber \\
 & =\frac{\sigma_{A}^{2}-\langle N_{A}\rangle}{\langle N_{A}\rangle^{2}}+\frac{\sigma_{B}^{2}-\langle N_{B}\rangle}{\langle N_{B}\rangle^{2}}-2\frac{C_{AB}}{\langle N_{A}\rangle\langle N_{B}\rangle}.\label{eq:nu_dyn}
\end{align}
 We see that $\nu_{dyn}$ will vanish for purely statistical fluctuation $\sigma^{2}=\langle N\rangle$ without inter-particle correlation $C_{AB}=0$. This observable has advantage of symmetry under the transposition of $A$ and $B$ and of independence of the detection efficiency. 

We firstly calculate the $\text{K}\pi$ fluctuations $\nu_{dyn,K\pi}$, where K refers to $\text{K}^{+}+\text{K}^{-}$ and $\pi$ refers to $\pi^{+}+\pi^{-}$. As discussed in Sec. \ref{sec:qfnc}, quark number fluctuation and correlation will obviously influence the variance and two-hadron correlations and therefore the $\nu_{dyn,K\pi}$. In Fig.~\ref{fig2}(a), we show $\nu_{dyn,K\pi}$ at different quark number fluctuation $\lambda_{1}$ and flavor conservation $\lambda_{2}$.  Decay contributions are also included. Here $\lambda_{1}\equiv\sigma_{f}^{2}/\langle N_{f}\rangle$ represents the variance of quark numbers with respect to the mean and we take the same value for $u$, $d$, and $s$ quarks. Pearson's correlation coefficient $\lambda_{2}=C_{f\bar{f}}/\sigma_{f}\sigma_{\bar{f}}$ is used to describe the flavor conservation of quark and antiquark and we also take the same value for $u\bar{u}$, $d\bar{d}$, and $s\bar{s}$ pairs. The averaged quark numbers are taken $\langle N_{u}\rangle=\langle N_{d}\rangle=\langle N_{\bar{u}}\rangle=\langle N_{\bar{d}}\rangle=1819$ and $\langle N_{s}\rangle=\langle N_{\bar{s}}\rangle=765$, which is corresponding to the size of quark system in the pseudo-rapidity interval $|\eta|\leq0.8$ in 0-5\% Pb-Pb collisions at $\sqrt{s_{NN}}$=2.76 TeV. We see that $\nu_{dyn,K\pi}$ increases with the increase of $\lambda_{1}$ and $\lambda_{2}$. The increase of quark number fluctuation $\lambda_{1}$ will increase the magnitudes of the first and second terms of $\nu_{dyn}$ in Eq.~(\ref{eq:nu_dyn}) by the variance of hadronic multiplicity, see Fig.~\ref{fig1}. The increase of $\lambda_{2}$ will mainly increase the hadronic pair correlation $C_{K^{+}K^{-}}$ and $C_{\pi^{+}\pi^{-}}$ and thus mainly increase the first and second terms of $\nu_{dyn,K\pi}$. Due to the similar reasons, results of $\nu_{dyn,Kp}$ and $\nu_{dyn,p\pi}$ also increase with $\lambda_{1}$ and $\lambda_{2}$, where $p$ refers to $p+\bar{p}$. In Fig.~\ref{fig2}(b), we show the result of $\nu_{dyn,K\pi}$ by direct combination (ini), that by including quark number fluctuation and flavor conservation in case of $\lambda_{1}=\lambda_{2}=1$, and that by further including decays, which illustrates these different contributions in final $\nu_{dyn,K\pi}$. 

The physical values of $\lambda_{1}$ and $\lambda_{2}$ in the context of central Pb-Pb collisions at ALICE detector are needed to discuss.  On the one hand, because the pseudo-rapidity coverage $|\eta|\leq0.8$ adopted by ALICE for $\nu_{dyn}$ measurements is only small fraction of the entire system ($y_{beam}>8$) created in heavy-ion collisions at LHC, $\lambda_{1}\sim1$ is reasonable in view of Poisson distribution in grand-canonical ensembles. On the other hand, pseudo-rapidity coverage $|\eta|\leq0.8$ is large for the conservation of quark flavor and we would expect $\lambda_{2}\sim1$, which is indicated from the observation that the radius of the measured charge balance is only about 0.45 \cite{AliceBF13} in central Pb-Pb collisions at LHC. The experimental datum of $\nu_{dyn,K\pi}$ \cite{alicepKpiFluct} in 0-5\% Pb-Pb collisions at $\sqrt{s_{NN}}=2.76$ TeV is shown in Fig.~ \ref{fig2}(a), where the center value of the datum is shown as a horizontal solid line and statistical and systematic uncertainties are shown as shadow bands, respectively. We indeed see that the results are close to the data as $\lambda_{1}$ and $\lambda_{2}$ take large values. However, we emphasize that such comparison is only served as a reference but not as a decisive test of QCM because the subsequent hadron re-scattering stage will also influence the dynamical fluctuation to a certain extent. 

\begin{center}
\includegraphics[width=0.9\linewidth]{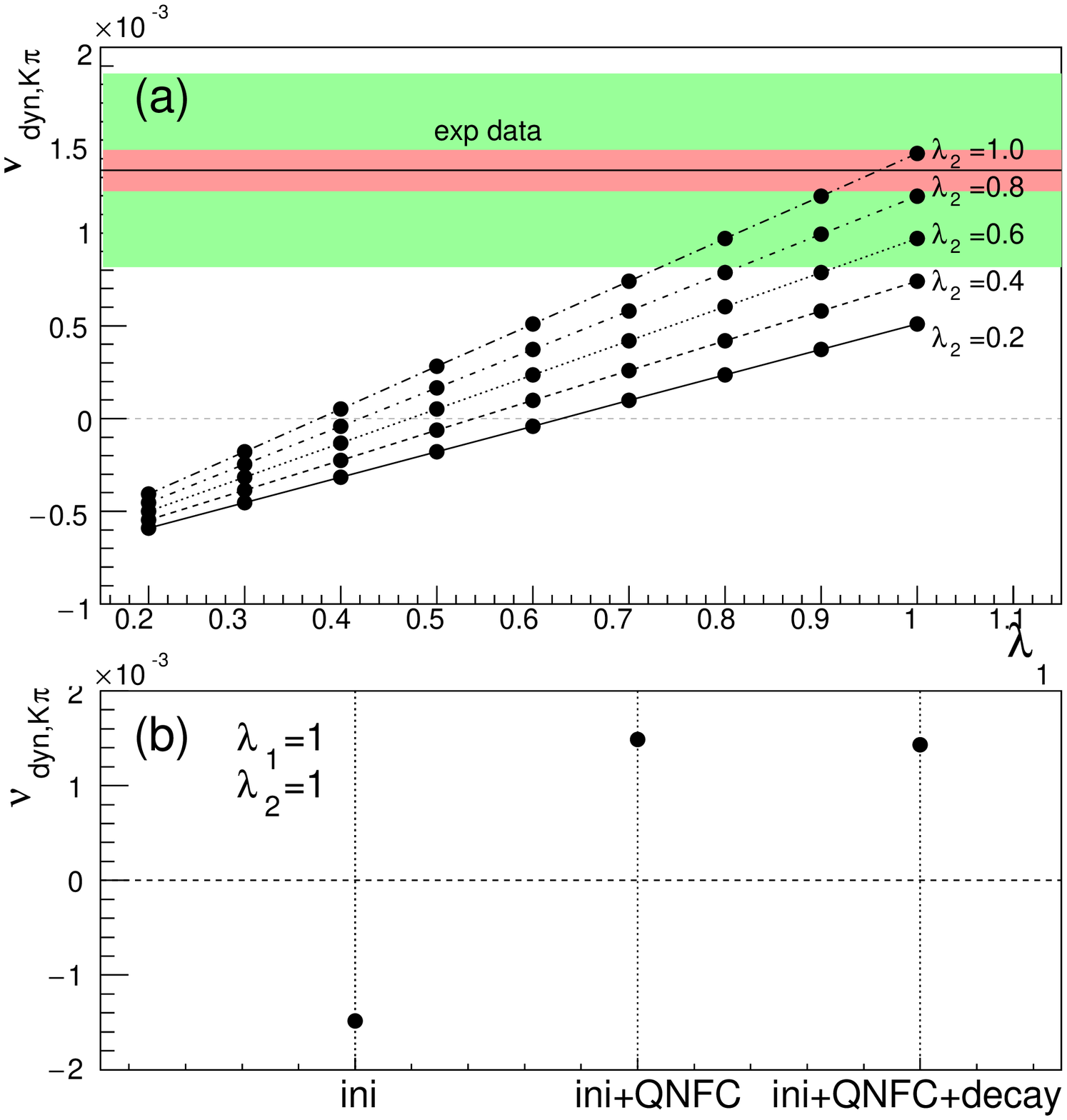}
\figcaption{\label{fig2}   (a) $\nu_{dyn,K\pi}$ at different values of quark number fluctuation $\lambda_{1}$ and flavor conservation $\lambda_{2}$;(b) result of $\nu_{dyn,K\pi}$ obtained by considering combination itself, that obtained by including quark number fluctuation and flavor conservation (QNFC) at $\lambda_{1}=\lambda_{2}=1$, and that obtained by further including the decays. The horizontal solid line with shadow bands in panel (a) is the datum of $\nu_{dyn,K\pi}$ in central Pb-Pb collisions at $\sqrt{s_{NN}}=2.76$ TeV \cite{alicepKpiFluct}. }
\end{center}

In Fig.~\ref{fig3}, we show the multiplicity dependence of $\nu_{dyn,K\pi}$ as well as those of $\nu_{dyn,Kp}$ and $\nu_{dyn,p\pi}$, where $p$ refers to $p+\bar{p}$. In QCM, hadronic variance $\sigma_{\alpha}^{2}$ and covariance $C_{\alpha\beta}$ are all proportional to the hadronic multiplicity, which can be seen from Eqs. (\ref{eq:c2b}), (\ref{eq:cBBbar}), (\ref{eq:c2m}), and (\ref{eq:cbm}). Therefore, following the definition Eq.~(\ref{eq:nu_dyn}) $\nu_{dyn}$ is inversely proportional to hadronic multiplicity and also $\langle dN_{ch}/d\eta\rangle$, and $\nu_{dyn}\times\langle dN_{ch}/d\eta\rangle$ for $K\pi$, $Kp$ and $p\pi$ are almost unchanged if $\lambda_{1}$ and $\lambda_{2}$ keep constant. The solid lines in Fig.~\ref{fig3} show results of QCM at fixed $\lambda_{1}=1$ and partial flavor conservation $\lambda_{2}=0.85$. The value of $\lambda_{2}$ is estimated by the measured charge balance function \cite{AliceBF13} via $\lambda_{2}\approx\int_{-0.8}^{0.8}B(\delta \eta)d\delta \eta/\int_{-\infty}^{\infty}B(\delta \eta)d\delta \eta$.  We notice that such inverse $\langle dN_{ch}/d\eta\rangle$ proportionality is one of main properties of $\nu_{dyn}$, which is also found in other models or event generators such as AMPT and HIJING, see the dashed lines and dotted lines in Fig.~\ref{fig3}, respectively, which are taken from Ref. \cite{alicepKpiFluct}. The experimental data of $\nu_{dyn,K\pi}$, $\nu_{dyn,Kp}$, and $\nu_{dyn,p\pi}$ are shown in Fig.~\ref{fig3}, which also exhibit such property in general, within the statistical and systematic uncertainties. 

The quantitative comparison among results of QCM, HIJING, AMPT and the experimental data is interesting and brings us some useful understanding.  First, our result of $\nu_{dyn,K\pi}$ (meson-meson pair) is slightly larger than that of HIJING, but those of $\nu_{dyn,pK}$ and $\nu_{dyn,p\pi}$ (baryon-meson pairs) are smaller than those of HIJING and are more close to the experimental data. Since the results of HIJING can be regarded as a superposition of independent $p-p$ collisions with string fragmentation hadronization and baryon production is more sensitive to the hadronization mechanism, the difference between our results and those of HIJING, to a certain extent, may be attributed to the creation of deconfined quark matter and its combination hadronization. Second, in comparison with HIJING, AMPT default version further includes the evolution of the hadronic re-scattering stage. Results of AMPT, dashed lines in Fig.~\ref{fig3}, are more close to the data, suggesting the nontrivial effects of hadronic re-scattering stage. If we make a naive estimation on the hadronic re-scattering effects by dividing the results of AMPT by those of HIJING to get a modification factor and multiplying it to results of QCM, we will see a significant improvement of the agreement with the ALICE data. 
\end{multicols}
\begin{center}
    \includegraphics[width=\linewidth]{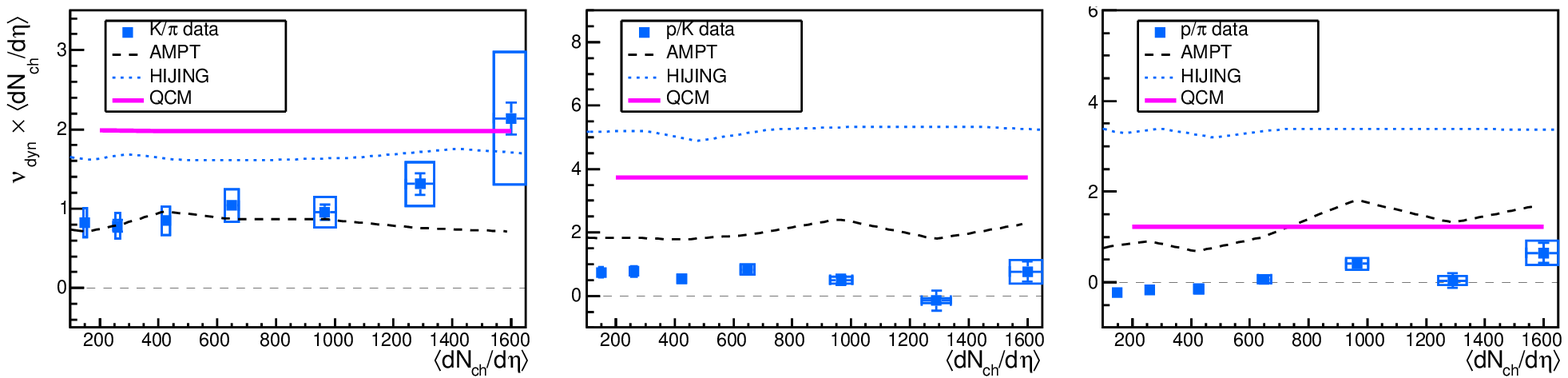}\figcaption{\label{fig3} Multiplicity dependence of dynamical fluctuations $\nu_{dyn}$ for $\text{K}\pi$, $\text{K}p$ and $p\pi$ pairs. The data in Pb-Pb collisions at $\sqrt{s_{NN}}=2.76$ TeV and results of AMPT and HIJING are taken from \cite{alicepKpiFluct}. }
\end{center}
\begin{multicols}{2}

\section{Summary}
In this paper, we have studied the second-order multiplicity fluctuation and correlation of identified mesons and baryons in the quark combination mechanism. We build a preliminary framework in which effects of different ingredient such as quark combination itself, quark number fluctuation and correlation at hadronization can be separated and specifically studied. Because fluctuation and correlation of hadrons in quark combination mechanism are mainly dependent on the constituent quark content of hadrons, we emphasize that there are lots of potentially interesting correlation properties among the results of different hadron species, which can be used to test the mechanism and, more importantly, obtain the information of fluctuation and correlation for the quark system at hadronization by virtue of the data of experimental observables.
As an example, we calculate a dynamical fluctuation observable $\nu_{dyn}$ for K$\pi$, $p\pi$ and K$p$ pairs in the context of Pb-Pb collisions at $\sqrt{s_{NN}}=2.76$ TeV, where we only consider the effects of the quark number fluctuation and the quark flavor conservation. 
In comparison with the experimental data and simulations of event generators, we find that the quark combination can reproduce the basic behavior of $\nu_{dyn}$ and hadronic re-scattering effects are not negligible.
In the upcoming work, we will systematically consider other nontrivial effects besides the hadronization such as those of hadronic re-scattering stage, finite acceptance, and nonzero baryon number density at lower energies, and will carry out a systemic comparison with the new $\nu_{dyn}$ data of LHC and also those of RHIC and make final predictions for more hadron species such as $p\Lambda$, K$\Lambda$, $p\Xi$ and $\Lambda\Xi$.

\vspace{-1mm}
\centerline{\rule{80mm}{0.1pt}}

\end{multicols}

\clearpage
\end{CJK*}

\vspace{2mm}


\begin{thebibliography}{90}

\vspace{3mm}

\bibitem{Shuryak80}E. V. Shuryak, Phys. Rept., 61, 71 (1980).%

\bibitem{qgp2004} see, e.g.,~Quark Gluon Plasma 3, edited by R.~C.~Hwa (Word Scientific, Singapore, 2004).

\bibitem{star12Moments} M. M. Aggarwal, \emph{et al.} (STAR Collaboration), Phys. Rev. Lett. \textbf{105}, 022302 (2010); L. Adamczyk, \emph{et al.} (STAR Collaboration), Phys. Rev. Lett. \textbf{112}, 032302 (2014).

\bibitem{kochFluct} V. Koch, A. Majumder, and J. Randrup, Phys. Rev.  Lett. \textbf{95}, 182301 (2005); M. Asakawa, U. V. Heinz, and B. Muller, Phys. Rev. Lett. \textbf{85}, 2072 (2000).

\bibitem{StephanovFluct} M. A. Stephanov, K. Rajagopal, and E. V. Shuryak, Phys. Rev. D \textbf{60}, 114028 (1999); M. A. Stephanov, Phys. Rev.  Lett. \textbf{102}, 032301 (2009).

\bibitem{Asakawa09} M. Asakawa, S. Ejiri, and M. Kitazawa, Phys.  Rev. Lett. \textbf{103}, 262301 (2009). %

\bibitem{Karsch11} F. Karsch and K. Redlich, Phys. Lett. B \textbf{695}, 136 (2011). 

\bibitem{NA492001} S. V. Afanasiev,\emph{ et al.} (NA49 Collaboration), Phys. Rev. Lett. 86, 1965 (2001).

\bibitem{Na492009} C. Alt, \emph{et al.} (NA49 Collaboration), Phys.  Rev. C \textbf{79}, 044910 (2009). 

\bibitem{Na492011} T. Anticic, \emph{et al.} (NA49 Collaboration), Phys. Rev. C \textbf{83}, 061902 (2011); Phys. Rev. C \textbf{87}, 024902 (2013).

\bibitem{star2009} B. I. Abelev, \emph{et al.} (STAR Collaboration), Phys.  Rev. Lett. \textbf{103}, 092301 (2009). 

\bibitem{star2015} N. M. Abdelwahab, \emph{et al.} (STAR Collaboration), Phys. Rev. C \textbf{92}, 021901 (2015). 

\bibitem{koch08} V. Koch, in Relativistic Heavy Ion Physics, edited by R. Stock (Springer, Heidelberg, 2010), pp. 626-652. 

\bibitem{fjh0912} J.~H. Fu, Phys. Lett. B \textbf{679} 209 (2009); Phys. Rev. C \textbf{85}, 064905 (2012).

\bibitem{tawfik} A. Tawfik, Prog. Theor. Phys. \textbf{126}, 279 (2011); Nucl. Phys. A \textbf{859}, 63 (2011); J. Phys. G \textbf{40}, 055109 (2013). 

\bibitem{Gorenstein2009} M. I. Gorenstein, M. Hauer, V. P. Konchakovski, E. L. Bratkovskaya, Phys. Rev. C \textbf{79}, 024907 (2009). 

\bibitem{Konchakovski2009} V. P. Konchakovski, M. Hauer, M. I. Gorenstein, E. L. Bratkovskaya, J. Phys. G \textbf{36}, 125106 (2009); J. Phys.  G \textbf{37}, 094045 (2010). 

\bibitem{koch2010} V. Koch, T. Schuster, Phys. Rev. C \textbf{81}, 034910 (2010). 

\bibitem{torr2007} G. Torrieri, Int. J. Mod. Phys. E \textbf{16}, 1783 (2007). 

\bibitem{koch13} A. Bzdak, V. Koch, V. Skokov, Phys. Rev. C \textbf{87},
014901 (2013). 

\bibitem{becattini02} F. Becattini and G. Passaleva, Eur. Phys. J.  C \textbf{23}, 551 (2002). 

\bibitem{shm03} See, e.g.,~P.~Braun-Munzinger, K.~Redlich and J.~Stachel, in Quark Gluon Plasma 3, edited by R.~C.~Hwa (Word Scientific, Singapore, 2004), pp.491-599. 

\bibitem{Andronc09} A. Andronic, P. Braun-Munzinger, J. Stachel, Phys. Lett. B 673, 142(2009). 

\bibitem{ShaoCN09}C. E. Shao, J. Song, F. L. Shao, and Q. B. Xie, Phys.\ Rev.\ C \textbf{80}, 014909 (2009).  

\bibitem{SJ13} J. Song, F. L. Shao, Phys. Rev. C \textbf{88}, 027901 (2013). 

\bibitem{Voloshin02} C. Pruneau, S. Gavin, and S. Voloshin , Phys.  Rev. C 66, 044904 (2002).

\bibitem{Christiansen09} P. Christiansen, E. Haslum, and E. Stenlund, Phys. Rev. C 80, 034903 (2009). 

\bibitem{alicepKpiFluct} M. Arslandok, for ALICE Collaboration, Nucl.  Phys. A 956, 870 (2016).

\bibitem{biro1999} T. S. Bir\'o, P. L\'evai, and J. Zim\'anyi, Phys. Rev.
C \textbf{59}, 1574 (1999). 

\bibitem{alcor2000} J. Zim\'anyi, T. S. Bir\'o,  T. Cs\"org\H{o}, and P.  L\'evai, Phys. Lett. B \textbf{472}, 243 (2000). 

\bibitem{Fries22003prl} R. J. Fries, B. M\"uller, C. Nonaka, and S.  A. Bass, Phys. Rev. Lett. \textbf{90}, 202303 (2003). 


\bibitem{ckm03} V. Greco, C. M. Ko, and P. Levai, Phys. Rev. C \textbf{68}, 034904 (2003).

\bibitem{hwa04} R. C. Hwa and C. B. Yang, Phys. Rev. C \textbf{70}, 024905 (2004).

\bibitem{co2006PRC} L. W. Chen and C. M. Ko, Phys.\ Rev.\ C \textbf{73}, 044903 (2006). 

\bibitem{sdqcm} F.~L.~Shao, Q.~B.~Xie and Q.~Wang, Phys.\ Rev.\ C \textbf{71}, 044903 (2005); 

\bibitem{wrq12} R. Q. Wang, F. L. Shao, J. Song, Q. B. Xie, Z. T.  Liang, Phys. Rev. C \textbf{86}, 054906 (2012). 

\bibitem{wrq15} R. Q. Wang, J. Song, F. L. Shao, Phys. Rev. C \textbf{91}, 014909 (2015). 

\bibitem{fries08review} R. J. Fries, V. Greco, P. Sorensen, Ann.  Rev. Nucl. Part. Sci. \textbf{58}, 177, (2008) and references therein. 

\bibitem{SJ12FB} J. Song, F. L. Shao, and Z. T. Liang, Phys. Rev.  C \textbf{86}, 064903 (2012). 

\bibitem{SFL17} F. L. Shao, G. J. Wang, R. Q. Wang, H. H. Li, and J. Song, Phys. Rev. C 95, 064911 (2017). 

\bibitem{SJ17fluct} J. Song, H. H. Li, R. Q. Wang, and F. L. Shao, Phys. Rev. C 95, 014901 (2017). 

\bibitem{Gavai05} R. V. Gavai and S. Gupta, Eur. Phys. J. C \textbf{43}, 31 (2005). 

\bibitem{Mukherjee06} S.~Mukherjee, Phys. Rev. D \textbf{74}, 054508 (2006). 

\bibitem{Dht15} H.~T. Ding, S.~Mukherjee, H. Ohno, P. Petreczky, and H. P. Schadler, Phys. Rev. D \textbf{92}, 074043 (2015). 

\bibitem{pdg2014} K. A. Olive, \emph{et al.} (Particle Data Group), Chin. Phys. C \textbf{38}, 090001 (2014).

\bibitem{AliceBF13} B. Abelev, \emph{et al.} (ALICE Collaboration), Phys. Lett. B \textbf{723}, 267 (2013). 



\end{thebibliography}
\end{document}